\documentclass{ws-ijmpa}
\usepackage[super,compress]{cite}
\usepackage{graphicx}

\begin{document}
\markboth{G.Gutierrez, M.A.Reyes}
{Fermilab Fixed Target Central Production}

%
\catchline{}{}{}{}{}
%

\title{FIXED TARGET EXPERIMENTS AT THE FERMILAB TEVATRON}

\author{Gaston Gutierrez}

\address{Fermi National Accelerator Laboratory\\
Batavia, Illinois, 60510, United States of America\\
gaston@fnal.gov}

\author{Marco A. Reyes}

\address{Physics Department, University of Guanajuato\\
Leon, Guanajuato, 37150, Mexico\\
marco@fisica.ugto.mx}

\maketitle

\begin{history}
\received{Day Month Year}
\revised{Day Month Year}
\end{history}

\begin{abstract}
This paper presents a review of the study of Exclusive Central Production at a Center of Mass energy of 
$\sqrt{s}=40$ GeV at the Fermilab Fixed Target program.  In all reactions reviewed in this paper, protons 
with an energy of 800 GeV were extracted from the Tevatron accelerator at Fermilab and directed to a 
Liquid Hydrogen target.  The states reviewed include $\pi^+\pi^-$, $K^0_s  K^0_s$, 
$ K^0_sK^\pm\pi^\mp$, $\phi\phi$ and $D^{*\pm}$.  Partial Wave Analysis results will be presented on the
light states but only the cross section will be reviewed in the diffractive production of $D^{*\pm}$.

\end{abstract}

\ccode{PACS numbers: 12.39.Mk, 14.40.Rt, 11.80.Et, 14.40.Be, 14.40.Lb, 13.85.Ni}


\section{Introduction}

The Double Pomeron Exchange (DPE) process was first observed at the CERN-ISR.\cite{isr}  Since then 
it is generally accepted that Central Production in hadron-hadron reactions at high center of mass and low 
momentum transfers proceeds through Double Pomeron Exchange.
It is also generally accepted that with its vacuum quantum numbers, the Pomeron is largely gluonic in nature.
Then proton-proton reactions, where the central object under study is well separated in rapidity from the 
protons, could be a good place to search for gluonium states.  In this article we review reactions of the type:
\begin{equation}
p_{beam}+p_{tgt} \to p_s \, (X) \, p_f 
\label{ppcol}
\end{equation}

\noindent
where the subscripts $s$ and $f$ refer to the slowest and fastest protons in the Laboratory reference frame. 

The Fixed Target (FT) program at the Fermilab Tevatron has produced a wealth of results,\cite{fnalexps}
but only experiment E690 used an 800 GeV/c proton beam on a Liquid Hydrogen target and had a beam
spectrometer capable of a precise measurement of the forward proton, making this the only experiment
in the Fermilab FT program that studied reactions of the type (\ref{ppcol}) above.  
In this article we review the main results of experiment E690 in the production of $\pi^+\pi^-$, $K^0_s  K^0_s$, 
$ K^0_sK^\pm\pi^\mp$, $\phi\phi$, and the diffractive production of $D^{*\pm}$.

\section{The E690 experiment}

\begin{figure}[ht]
\centerline{\includegraphics[width=8cm]{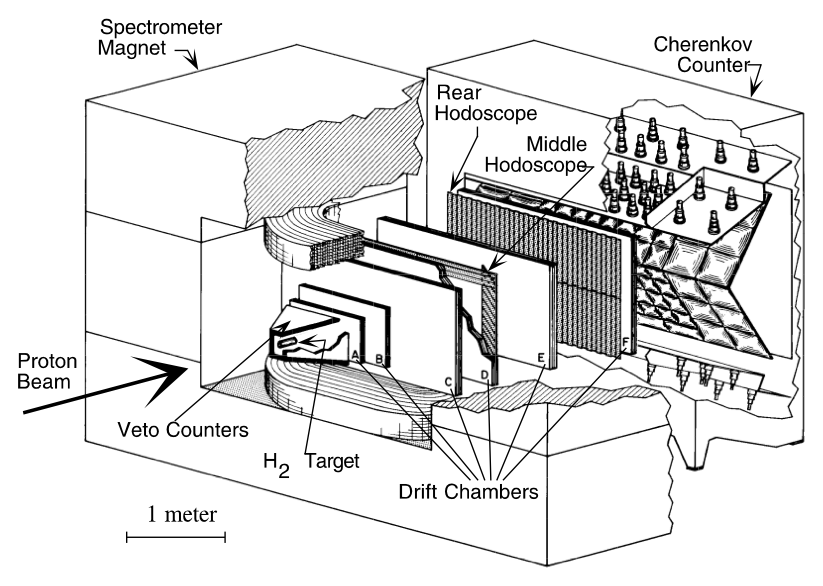}}
\caption{The E690 main Spectrometer. \label{jgg}}
\end{figure}

The E690 experiment consisted of an 800 GeV/c proton beam hitting a liquid hydrogen (LH$_2$) target.  
A high rate, open geometry multiparticle spectrometer, shown in Fig. \ref{jgg}, followed the hydrogen target.  
A beam spectrometer was used to accurately measure the 800 GeV/c beam and the scattered proton.  
Details of the spectrometer can be found in Ref. \citen{edh}.

\noindent
All final states reviewed in this article were subject to the following event selection:

\begin{itemize} 
\item a primary interaction vertex within the fiducial region of the LH$_2$ target,
\item a number of charged tracks consistent with the desired topology,
\item a number of secondary vertexes also consistent with the desired topology,
\item a fast proton, $p_f$, measured in the forward spectrometer.
\end{itemize}

\noindent
For Central Production and low $p_t$ the slow proton $p_s$ often stays within the LH$_2$ target.   
Therefore E690 required that $p_s$ would not be reconstructed in the main spectrometer and the missing 
mass squared of the event ($M^2_{miss}$) was used to identify this proton by requiring that $M^2_{miss}$ would
agree with the mass squared of the proton. 
Other selection criteria pertinent to particular final states are discussed in each section below.


\section{Central production of the $\pi^+\pi^-$ system}

\begin{figure}[ht]
\centerline{\includegraphics[height=3.9cm]{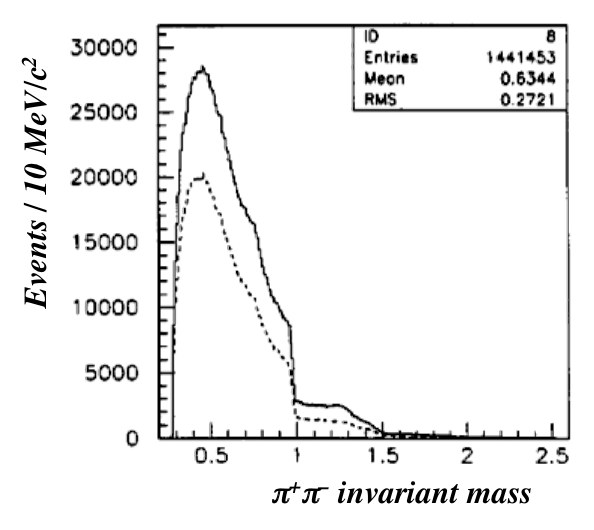}
\includegraphics[height=3.9cm]{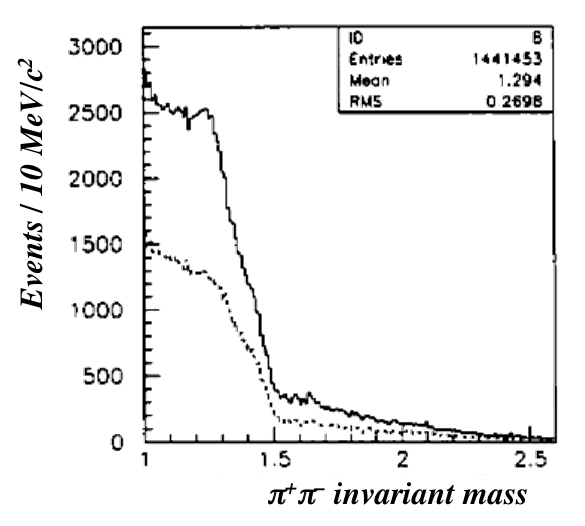}
\includegraphics[height=3.9cm]{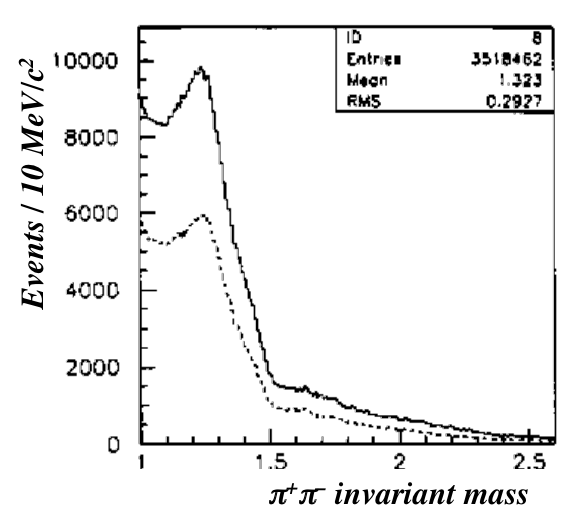}}
\caption{Left panel: $\pi^+\pi^-$ invariant mass distribution for events with $p_{t,s}^2<0.1 $(GeV/c)$^2$ 
(solid line) and with both $p_{t,s}$ and $p_{t,f}^2<0.1 $(GeV/c)$^2$ (dashed line).  
Center panel: same plot as in the left panel but with a different mass scale.  
Right panel: same $\pi^+\pi^-$ invariant mass distribution without the selection in $p_{t,f}$
to show the presence of the $f_2(1270)$ resonance; the dashed line includes the $x_F$ selection the solid
line does not.}
\label{f1}
\end{figure}

In order to select events according to reaction \ref{ppcol} with $X\to\pi^+\pi^-$, experiment E690 required 
\cite{pipi,Kyriacos}, in addition to the protons, two charged tracks attached to the primary vertex, both of them 
with \v{C}erenkov identities compatible with being pions; the missing proton longitudinal momentum was required 
to be $|p_l|<1.0$GeV/c, and a rapidity difference between the missing proton and either pion greater than 1.8 
rapidity units to avoid $\Delta^{++}$ contamination.  To select centrally produced events, the Feynman $x_F$ of 
the $X$ system was required to be between $-0.1<x_F<0.0$.  The transverse momentum of both scattered protons 
was required to be $p_t^2<0.1$(GeV/c)$^2$, to enhance $S$-wave dominance in the selected events.  
The $\pi^+\pi^-$ invariant mass distribution for the selected events is shown in Fig.~\ref{f1}.

A Partial Wave Analysis (PWA) was performed following the method described by Chung and Trueman\cite{chung}.  The
reflectivity basis was used for the analysis, with eigenvectors defined in the Gottfried-Jackson (GJ) frame.  In 
the rest frame of the $X$ system, E690 defined the GJ frame with the $z$-axis in the direction of the momentum 
transfer of the beam proton, with the $y$-axis perpendicular to the plane defined by the momentum transfers in 
the overall centre of mass (CM), and the $x$-axis defined as in a right handed coordinate system.
Only the amplitudes with $l\leq 2$ and $m\leq 1$ were used in the analysis:


\begin{eqnarray}	
\begin{array}{lll}
S_0^- = Y_0^0 , \;\;\; &  P_0^- = Y_1^0 , \;\;\; &  D_0^- = Y_2^0 \;\;\; \\[8pt]
& P_1^\pm = \left( Y_1^1 \right. \pm \left. Y_1^{-1} \right)/\sqrt{2} , \;\;\; &
  D_1^\pm = \left( Y_2^1 \right. \pm \left.Y_2^{-1} \right)/\sqrt{2} \;\;\; \\[8pt]
\end{array}
\label{waves}
\end{eqnarray}

\begin{figure}[ht]
\centerline{\includegraphics[height=4.5cm]{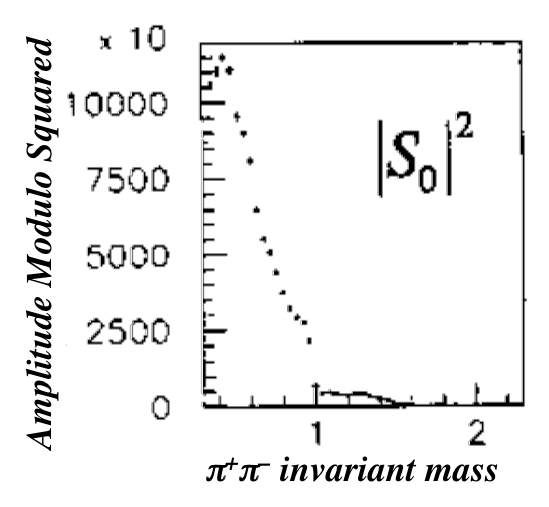}
\includegraphics[height=4.5cm]{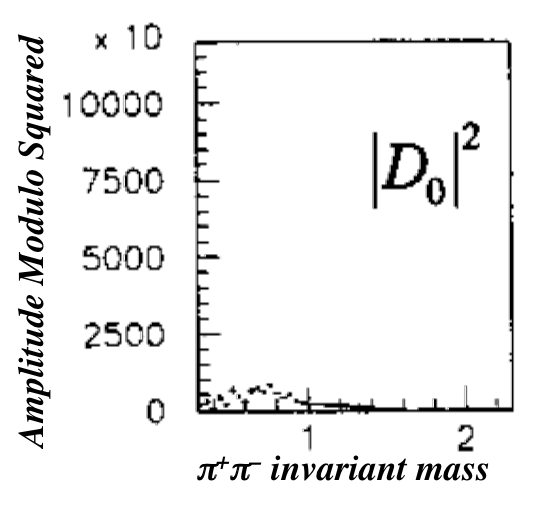}}
\caption{$|S|^2$ and $|D_0|^2$ wave intensities as a function of the $\pi^+\pi^-$ invariant mass measured 
in the final PWA analysis.}
\label{f2}
\end{figure}

The assumption of $S$-wave dominance near threshold is sufficient to select a single, continuous solution 
throughout the considered mass spectrum, from threshold to 1.5 GeV/c$^2$, for events with both 
$p_T^2<0.1$ (GeV/c)$^2$.  The $S_0$ and $D_0$ waves are plotted in Fig.\ref{f2}.  The $D_+$-wave contribution 
is about the same as the $D_0$ one, and the $D_-$-wave contribution is essentially zero. All $P$-wave 
contributions are negligible, as expected from double Pomeron exchange.
When the low transverse momentum selection on the fast proton is removed a significant $D$-wave contribution 
is observed above 1 GeV/c$^2$ due to the production of the $f_2(1270)$ (see Right panel in Figure \ref{f1}).

The dotted lines on the left and center plots in Figure \ref{f1} and the data points on the left plot in
Figure \ref{f2} show two prominent drops in the $\pi^+ \pi^-$ invariant mass spectrum.  
The sharp drop at 1 GeV was first explained by Morgan and Pennington \cite{AMP} as the interference
of the $f_0(980)$ with a background that has a phase of about 90 degrees at 1 GeV.  The second drop at 1.5 GeV
is due to the interference of the same background with the $f_0(1500)$.  These same interference effects are 
observed in $\pi \pi$ elastic scattering (for a review see the $\pi \pi$ scattering section in 
W. Ochs\cite{ochs}).  
No other features are observed in the $S$-wave spectrum, for example there is essentially no evidence for the 
$f_0(1370)$ or the $f_0(1710)$.



\section{Central production of the $ K^0_s K^0_s$ system}

The event selection for the decay $X\to  K^0_s K^0_s$ included two secondary vertices with a tight $ K^0_s$ 
invariant mass.  The background under the $ K^0_s$ invariant mass peak was so small that no direct particle
identification was needed.\cite{ksks} 
For every event the difference in rapidity between $p_s$ and the $ K^0_s K^0_s$ system was required to be 
larger than 1.2 units.  For $p_f$ this difference was larger that 3.7 units.

\begin{figure}[ht]
\centerline{\includegraphics[height=4.5cm]{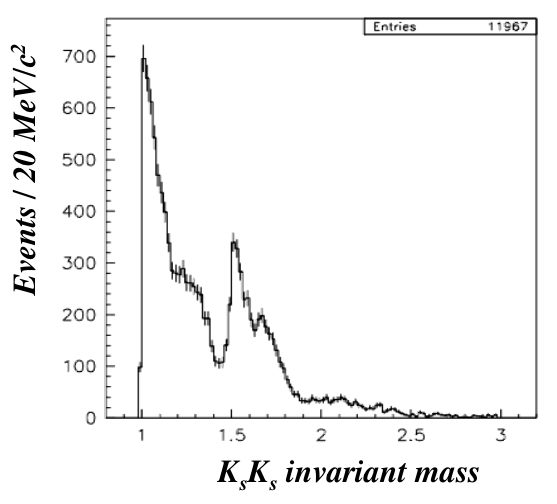}
\includegraphics[height=4.5cm]{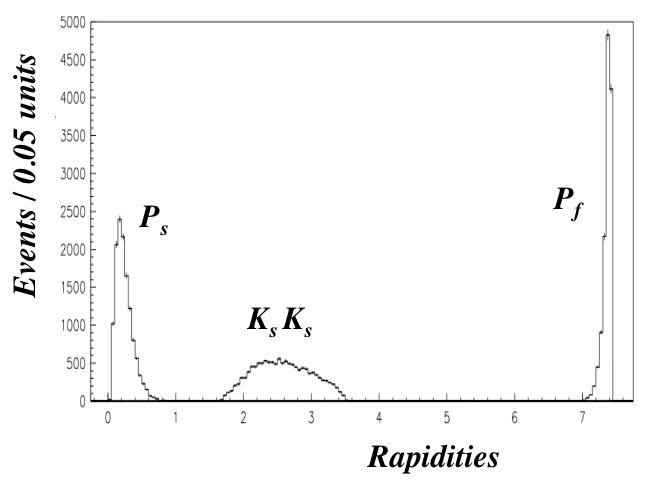}}
\caption{Left panel: $ K^0_s K^0_s$ invariant mass distribution after the final event selection.  
Right panel: Rapidity distributions for the slow proton, the $ K^0_s K^0_s$ system, and the fast proton.}
\label{f3}
\end{figure}

The left panel in Figure \ref{f3} shows the $ K^0_s K^0_s$ invariant mass between threshold and 3 GeV/c$^2$.
This mass distribution is smooth beyond 2 GeV/c$^2$, with no evidence of the narrow $f_J(2220)$ state seen by 
the BES Collaboration.\cite{bes,pdg96}  The right panel in the same figure shows the rapidity distributions
for $p_s$, $ K^0_s K^0_s$ and $p_f$.

A PWA analysis of the 11182 selected events was performed in bins of the $ K^0_s K^0_s$
invariant mass for events in the range $-0.22\leq x_F(X) \leq -0.02$, integrating over $p_{t,s}^2$, 
$p_{t,f}^2$, and $\delta$, the angle between the scattered protons in the $ K^0_s K^0_s$ center of mass.  
Using the reflectivity basis for this parity conserving system, the wave amplitudes used were 
$ S_0^-$, $ D_0^-$, $ D_1^-$ and $ D_1^+$ (see Eq. \ref{waves}).  Only (even)$^{++}$ waves are allowed in 
the $ K^0_s  K^0_s$ system.\cite{chung2}  E690 performed the partial wave analysis maximizing the extended 
likelihood with respect to the four wave moduli and the two relative phases $\phi(D_0^-)-\phi(S_0^-)$ and 
$\phi(D_1^-)-\phi(S_0^-)$. 

\begin{figure}[ht]
\centerline{\includegraphics[height=7.5cm]{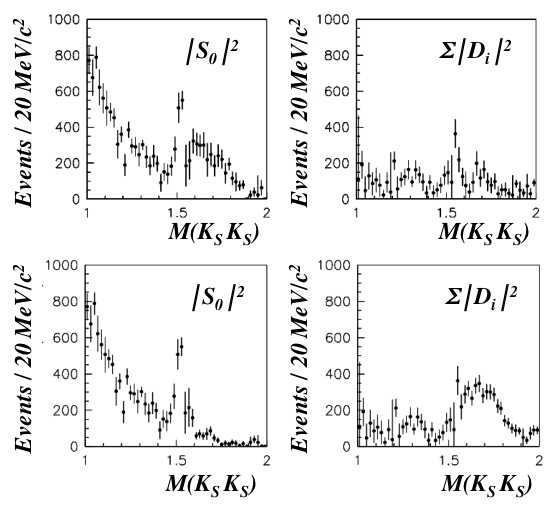}}
\caption{Results of the PWA analysis on the $ K^0_s K^0_s$ system.  The plots on the left (right) show $|S|^2$ 
($|D|^2$) as a function of the $ K^0_s K^0_s$ invariant mass.  As explained in the text there is a bifurcation
point at 1.55 GeV/c$^2$ giving rise to two solutions.  The upper (lower) plots show the solution with the
largest contribution of $S$ ($D$) wave after the bifurcation point.}
\label{f4}
\end{figure}

Using the above four waves there are two solutions for every mass bin.  These solutions can be continued
from one bin to the next as long as they do not cross. This problem can be expressed in terms
of Barrelet zeros\cite{barr}, when a zero crosses the real axis the solution bifurcates.  In the E690 analysis
one of the two Barrelet zeros becomes real at 1.55 GeV/c$^2$ producing a bifurcation point.  
At threshold the $ K^0_s K^0_s$ cross section is dominated by the $f_0(980)$ resonance, so before the 
bifurcation point E690 eliminated the solution with a small contribution of $S$ wave near threshold.  Of the two
solutions after the bifurcation point, one is mainly $S$ wave and the other mainly $D$ wave.  
These two solutions are shown in Figure \ref{f4}.

Three main features are observed before the bifurcation point: ({\it i}) the well established $f_0(1500)$
is clearly seen,  ({\it ii}) the $f_2(1270)$ is observed in the $D$ wave amplitude, and ({\it iii}) there
is no evidence of the $f_0(1370)$.  After the bifurcation point E690 could not determine the spin of the 
so-called $f_J(1710)$.  In a similar PWA analysis in $K^+ K^-$ and $K^0_s K^0_s$, WA102 later favored 
the spin-0 interpretation of the $f_J(1710)$.\cite{wa102}


The classification of the scalar mesons has not yet been resolved.
The most commonly accepted interpretation is that there are two $q\bar q$ meson resonances in 
the region between 1300 and 1900 GeV/c$^2$, and that the three observed states $f_0(1370)$, $f_0(1500)$ 
and $f_0(1710)$ are a mixture of those two $q\bar q$ states and the lowest mass scalar glueball.\cite{ochs,wa102}  
However, E690 did not find evidence of a scalar resonance in the region of the $f_0(1370)$ in either 
the $\pi^+\pi^-$ or the $ K^0_s K^0_s$ systems. This result is in agreement with the absence of the $f_0(1370)$
in $\pi \pi$ elastic scattering.\cite{ochs}



\section{Central production of the $ K^0_sK^\pm\pi^\mp$ system}

In order to study reaction \ref{ppcol} with $X\to  K^0_sK^\pm\pi^\mp$, experiment E690 selected events 
where the central 
cluster consisted of one positive track, one negative track, and a $ K^0_s$.\cite{kkpi}  At least one of the 
two charged tracks was required to be identified by the \v{C}erenkov counter as either a $\pi$, or an 
ambiguous $K/p$, and the other track was required to have an identity compatible with the final state. In 
all selected events, the forward proton, $p_f$, was separated from the central mesons by at least 3.5 
units of rapidity.   A minimum gap of 1.8 units of rapidity was required between each individual meson and 
$p_s$ to ensure that there was no contamination of the final state from reactions in which $p_s$ would be a 
decay product of a baryon resonance, such as $\Delta^{++}$ or $\Lambda(1520)$. Finally, in order to ensure 
near uniform acceptance, the $x_F$ of the meson system was required to be in the range [$-$0.15, $-$0.02͔].

\begin{figure}[ht]
\centerline{\includegraphics[height=4.5cm]{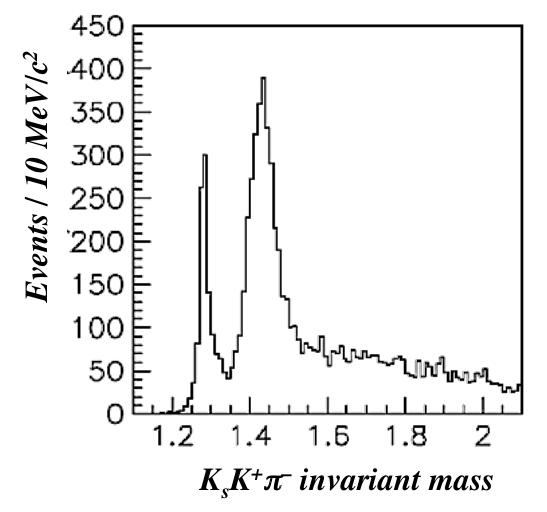}
\includegraphics[height=4.5cm]{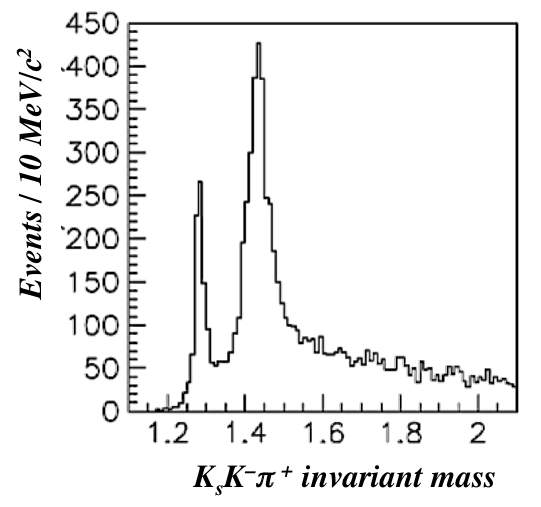}}
\caption{$ K^0_sK^+\pi^-$ and $ K^0_sK^-\pi^+$ invariant mass distributions after final event selection.}
\label{f5}
\end{figure}

The $K\bar K\pi$ invariant mass distribution for both charge states is shown in (Fig.~\ref{f5}).  The first
peak is easily identified by its mass and width as the $f_1(1285)$, and the second peak is nowadays 
identified as the $f_1(1420)$ meson.  At the time of the E690 publication\cite{kkpi} there 
were disagreements as to whether this second peak corresponded to the $f_1(1420)$ 
state,\cite{armstr,barberis,aihara} or to a $0^{-+}$ state decaying to $a_0 \pi$ that had been seen in $K^-p$ 
interactions\cite{f1-1510}. This ambiguity was known as the {\it $E/\iota$ puzzle.}\cite{eiota}

\begin{figure}[ht]
\centerline{\includegraphics[height=6.3cm]{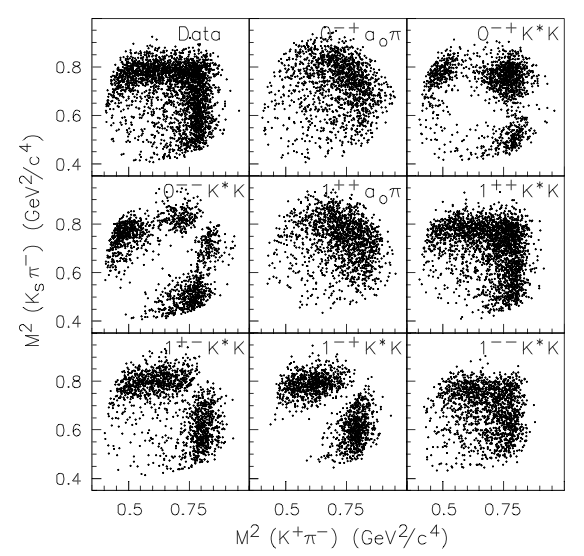}
\includegraphics[height=6.3cm]{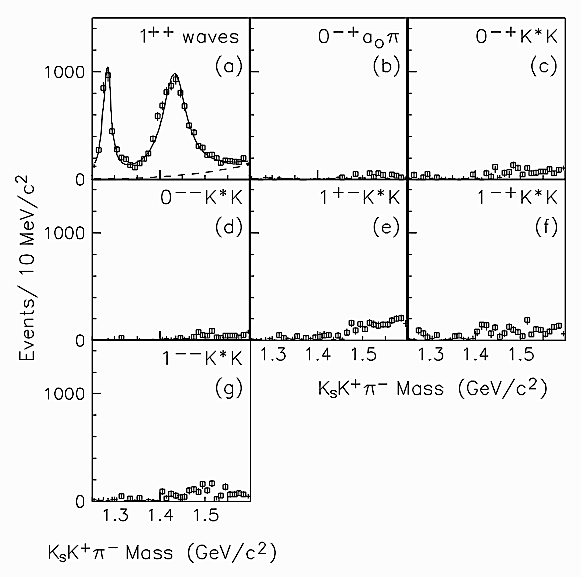}}
\caption{Left panel: $ K^0_sK^+\pi^-$ Dalitz plots for the data and several MC waves in the $f_1(1420)$ region.
Right panel: results of the PWA analysis on the $ K^0_sK^+\pi^-$ final state.  The other charge state gives
the same results.\cite{kkpi}}
\label{f6}
\end{figure}

The Dalitz plots in the $f_1(1420)$ region are shown in the left panel of Figure \ref{f6} for the data, and the 
Monte Carlo for several different waves.  Just by inspection of these plots it is easy to see that the data 
matches the $1^{++} K^*K$ simulation very well. To assess if there are small amounts of other waves, E690 
performed a PWA analysis between threshold and 1.6 GeV.\cite{kkpi}  The results, shown in the right panel in 
Figure \ref{f6}, clearly demonstrated that only $1^{++}$ waves were needed to describe the data, 
confirming that pseudoscalar states were not seen in central production,\cite{armstr} and solving 
the {\it $E/\iota$ puzzle.}


\section{Central production of the $\phi\phi$ system}

The first observation of the OZI \cite{ozi} suppressed reaction $\pi^-p\to \phi\phi n$, was made using 
the BNL$-$MPS spectrometer.\cite{mps} A subsequent PWA analysis on a larger data sample showed that 
three $2^{++}$ states were necessary to fit the data.\cite{etkin}  Fermilab FT experiment E623 measured the
$\phi\phi$ cross section in the reaction $p N \to \phi\phi X$ but did not perform a PWA analysis.\cite{Fphiphi}
Given the OZI suppression the $\phi\phi$ channel is believed to be a good place to look for the production of
glueballs.

\begin{figure}[ht]
\centerline{\includegraphics[height=4.5cm]{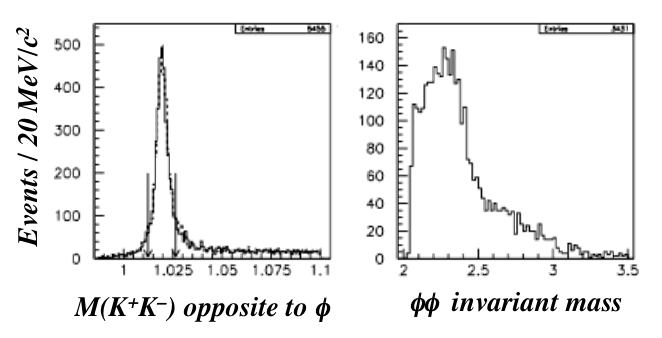}}
\caption{Left panel: The $K^+K^-$ invariant mass when the other $K^+K^-$ pair lay in the $\phi$ mass region.  
Right panel: $\phi\phi$ invariant mass distributions for the selected events.}
\label{fifim}
\end{figure}

E690 measured the Central Production of $\phi\phi$ in reaction \ref{ppcol} with $X\to\phi\phi$
and $\phi \to K^+K^-$,\cite{phiphi}
selecting events with four charged tracks coming from the primary vertex, in addition to the diffracted proton. 
The tracks were required to have \v{C}erenkov information compatible with being kaons, and at least 
one of them identified as not being a pion.  A kinematic cut on the missing momentum $p_z<250\,$MeV 
or arctan($p_t/p_z)>$45 degrees ensured that the missing proton was outside of the detector's 
geometrical acceptance.  The $\phi$ mass region was defined as $1.0124<m(K^+K^-) <1.0264$ GeV/c$^2$.  
The $K^+K^-$ and $\phi\phi$ invariant mass distributions for the selected events are shown in Fig. \ref{fifim}.

E690 performed a PWA analysis of the $\phi\phi$ system using states defined in terms of the total angular 
momentum $J$, orbital angular momentum $L$, parity $P$ and exchange reflectivity $\eta$:
\begin{equation}
G^{J^P LSM^\eta}(\gamma,\beta,\alpha_1,\alpha_2,\theta_1,\theta_2) =
\mbox{Real{\Huge [}} \frac{(1-i)-\eta(1+i)}{2} 
\sum_{\mu,\lambda} C(1,1,S|\mu,-\lambda) \times \nonumber
\end{equation}
\vspace*{-7mm}
\begin{equation}
c(l,s,j|0,\mu-\lambda) e^{-iM\gamma} e^{i\mu\alpha_1} e^{i\lambda\alpha_2}
d_{M,\mu-\lambda}^J(\beta) d_{\mu,0}^1(\theta_1) d_{\lambda,0}^1(\theta_2) 
\mbox{\Huge ] } \label{gjplsm}
\end{equation}

\noindent
where $M=|J_z|$.  $\gamma$ and $\beta$ are defined as the GJ angles of one of the $\phi$ mesons in 
the rest frame of the $\phi\phi$ system, with the $z$-axis in the direction of $\vec p_{fast}-\vec p_{beam}$, 
and the $y$-axis in the direction of the $(\vec p_{fast}-\vec p_{beam})\times (\vec p_{slow}-\vec p_{tgt})$ 
cross product, measured in the $pp$ CM system.  The other angles ($\alpha_{1,2},\theta_{1,2}$) are the two 
pairs of GJ angles of the $K^+$'s in their parent $\phi$ rest frames, with the $z'$-axis in the direction 
of $\vec p_\phi$, and $y'=z\times z'$.  This system has $I=0$, $C=+$, and $L+S=$even number.

\begin{figure}[ht]
\centerline{\includegraphics[width=13.0cm]{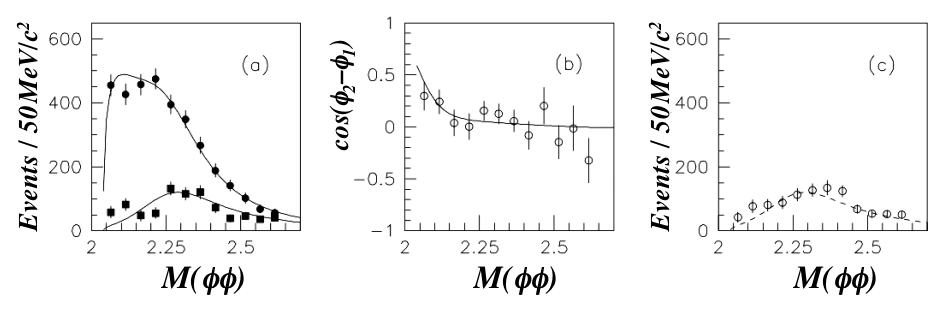}}
\caption{The symbols with error bars show three $J^{PC}LS=2^{++}02$ waves needed in the PWA analysis of the
$\phi\phi$ system. The upper/lower points in (a) correspond to the $M^\eta=0^-/1^-$ amplitudes. 
The phase between those waves is shown in (b). Plot (c) shows the $M^\eta=1^+$ amplitude. 
The lines represent the fit to the data described in the text.}
\label{fifiw}
\end{figure}

The analysis was performed in twelve bins of 50 MeV/c$^2$ between 2.04 and 2.64~GeV/c$^2$.  
The result of the analysis is shown by the symbols with error bars in Figure~\ref{fifiw}.  Only three waves 
were necessary to describe the data.  All waves have quantum numbers $J^{PC}LS=2^{++}02$, with $M^\eta=0^-/1^-$ 
for the upper/lower symbols in Fig.~\ref{fifiw}.a, and $M^\eta=1^+$ for the circles in Fig.~\ref{fifiw}.c.  
The phase between the interfering waves $M^\eta=0^-/1^-$ is shown in Fig.~\ref{fifiw}.b.

The $\phi\phi$ cross section opens at threshold very fast, reminiscent of the fast opening of the $KK$ cross 
section due to the presence on the $f_0(980)$ just below threshold.  This can be taken as an indication of a 
state right below threshold waiting to go into $\phi\phi$.  E690 performed a fit to the result of the PWA analysis
using a resonance below threshold produced as $M^\eta=0^-$ and another resonance above threshold produced in
all $M^\eta$ states.  The results of the fit are shown by the lines in Fig. \ref{fifiw}. The extracted
parameters for the resonance above threshold are:

\begin{eqnarray}
\begin{array}{rl}
M_R&=2.243 \pm 0.015(stat) \pm 0.010(syst) \;\; \mbox{ GeV}/c^2 \\
\Gamma_R&=0.368 \pm 0.033(stat) \pm 0.030(syst) \;\; \mbox{ GeV}/c^2
\end{array}
\end{eqnarray}

\noindent
With the available statistics, and only fitting to the $\phi\phi$ channel, the parameters for the resonance 
below threshold could only be determined approximately, giving $M_R \sim 1.9$ GeV/c$^2$ and 
$\Gamma_R \sim 0.3$ GeV/c$^2$.  These values are consistent with either the $f_2(1950)$ or the 
$f_2(2010)$.\cite{pdg}



\section{Diffractive production of charm}

To search for intrinsic charm states within the proton that could be excited diffractively,\cite{brodsky} 
E690 selected inclusive events with a $D^*$ meson decaying to $K\pi\pi$:\cite{wang}
\begin{equation}
pp\to Y \left[ D^*\to(D^0\to K\pi)\pi \right] p_f
\label{charmdec}
\end{equation}
with $Y$ being an unidentified recoil system.  Even though the $D^*$ is centrally produced and the 
forward proton is clearly diffractive this reaction does not qualify as exclusive, but we decided to include 
it in this review because it constitutes the first measurement of the diffractive charm cross section.

\begin{figure}[ht]
\centerline{\includegraphics[height=3.8cm]{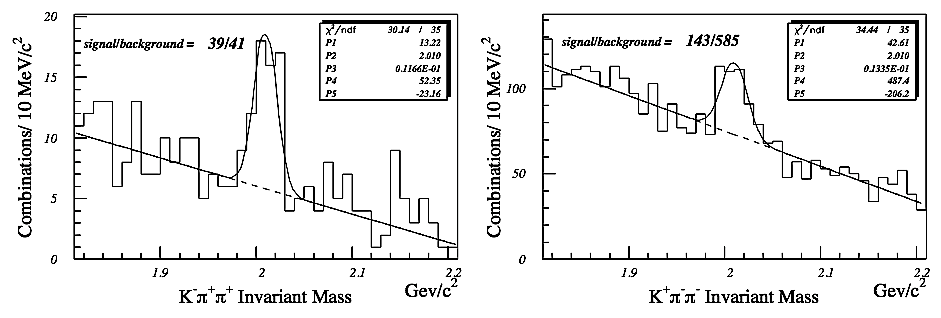}}
\caption{$K\pi\pi$ invariant mass distributions for the selected events in reaction (\ref{charmdec}).
The panel on the left shows $D^{*+}$ and the right one $D^{*-}$.  The lines are fits to a Gaussian plus a 
linear background.}
\label{f7}
\end{figure}

The events were selected requiring at least four charged tracks (including the scattered beam proton) with the
correct charge assignments coming from the primary vertex.  
The two tracks from the $D^0$ decay were required to be identified by the \v{C}erenkov counter as a kaon 
and a pion, the slow $\pi^+$ from the $D^{*+}$ decay was identified by the time-of-flight system.  The identification
of the slow $\pi^-$ from the $D^{*-}$ decay was not required since this background is composed mostly of $\pi^-$.
The $D^*$ was selected requiring that $|Q-5.83|<0.5$MeV/c$^2$, where $Q=M(K\pi\pi)-M(K\pi)-M(\pi)$.  The $x_F$
of the diffractive proton was required to be larger than $x_F>0.85$.\cite{goul}
Figure \ref{f7} shows the $K^-\pi^+\pi^+$ and the $K^+\pi^-\pi^-$ mass distributions for the selected events.

\begin{figure}[ht]
\centerline{\includegraphics[height=3.8cm]{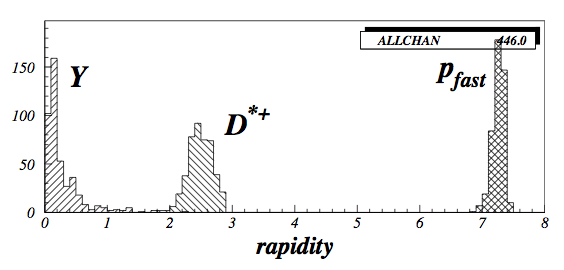}}
\caption{$D^*$, $Y$ and $p_f$ rapidity plots for the $D^{*+} \to D^0 (K^− \pi^+ )\pi^+$ 
decay in reaction (\ref{charmdec}).}
\label{f8}
\end{figure}

Figure \ref{f8} shows the rapidity distribution for the $D^{*+}$, the diffractive proton $p_f$ and the recoil
system $Y$.\cite{wang2}  We can see a clear gap between the central $D^{*+}$ and the rest of the system.

The values of the measured $D^*$ diffractive cross sections are:

\begin{eqnarray}
\begin{array}{rl}
\sigma_{\mbox{diff}}(D^{*+}) &= 0.185 \pm 0.044 (\mbox{stat}) \pm 0.054 (\mbox{syst}) \;\; \mu b \\
\sigma_{\mbox{diff}}(D^{*-}) &= 0.174 \pm 0.034 (\mbox{stat}) \pm 0.029 (\mbox{syst}) \;\; \mu b
\end{array}
\end{eqnarray}

These cross sections are compatible with previous limits and predict a total charm diffractive cross section
of $\sigma_{\mbox{diff}}(c\bar c) \sim$ 0.66 $\mu$b.  This gives a ratio of the diffractive charm to the total
$pp$ diffractive cross section of $\sim 10^{-4}$, which is 10 times smaller than the ratio of the inelastic
charm to the inelastic $pp$ cross section.\cite{wang}


\section{Conclusions}

In this paper we reviewed the Exclusive Central Production of $\pi^+ \pi^-$, $K^0_s K^0_s$, 
$K^0_s K^\pm \pi^\mp$ and $\phi\phi$ and the diffractive production of charm at the Fermilab Fixed 
Target program.  All these reactions were produced with an 800 GeV/c proton beam hitting a LH$_2$
target.  A Partial Wave Analysis was performed in all light meson production reactions, only the 
cross section was measured in the diffractive production of charm.
The scalar mesons $f_0(980)$ and $f_0(1500)$ are clearly seen both in $\pi^+ \pi^-$
and $K^0_s K^0_s$.  There is no clear evidence of the $f_0(1370)$ in either reaction, a result that is
in agreement with $\pi^-\pi^+$ elastic scattering.  The spin of the $f_J(1710)$ could not be determined
with this data alone.  Only two resonances, the $f_1(1285)$ and the $f_1(1420)$, are seen in the PWA analysis 
of $K^0_s K^\pm \pi^\mp$.  The existence of a $0^{-+}$ state decaying to $a_0 \pi$ in this reaction is
completely ruled out.  The Central Production of $\phi\phi$ is clearly seen.  Two $2^{++}$ resonances, 
one below and another one above threshold, are needed to explain the $\phi\phi$ data.  The measurement of 
the $D^*$ diffractive cross section shows that the ratio of the diffractive charm to the diffractive $pp$ 
cross section is ten times smaller than the ratio of the inelastic charm to the inelastic $pp$ cross section.


\section*{Acknowledgments}

We would like to thank the many people that provided information about the Fermilab 
Fixed Target program.  This work was funded by the US Department of Energy and CONACYT Mexico.
M. A. Reyes would like to thank CONACYT for supporting a sabbatical stay at Femilab while this paper
was being written.

\end{document}